\documentclass[12pt]{iopart}
\usepackage{amsfonts}
\usepackage{amssymb}
\usepackage[dvips]{graphicx}
\usepackage{colordvi}
\usepackage{times}

\baselinestretch

\newcommand{\be}{\begin{equation}}
\newcommand{\ee}{\end{equation}}
\newcommand{\ran}{\right\rangle}
\newcommand{\lan}{\left\langle}

\newcommand{\id}{\textrm{d}}

\newcommand{\bea}{\begin{eqnarray}}
\newcommand{\eea}{\end{eqnarray}}

\begin{document}

\title{Elements of a unified framework for response formulae}

\author{Matteo Colangeli$^1$, Valerio Lucarini$^{2,3}$}
\address{$^1$ Dipartimento di Matematica, Politecnico di
Torino, Corso Duca degli Abruzzi 24, 10129 Torino, Italy\\
$^2$ Meteorologisches Institut, University of Hamburg\\KlimaCampus, Grindelberg 7,
20144 Hamburg, Germany\\
$^3$ Department of Mathematics and Statistics\\University of Reading, Reading, RG6
6AX, UK}

\ead{colangeli@calvino.polito.it, valerio.lucarini@zmaw.de}

\begin{abstract}
We provide a physical interpretation of the first and second order terms occurring in Ruelle's response formalism. We show that entropy fluxes play a major role in determining the response of the system to perturbations. Along this line, we show that our framework allows one to recover a wealth of previous results of response theory in both deterministic and stochastic contexts. In particular, we are able to shed light on the crosslinks between
the dynamical systems approach \'{a} la Ruelle and large deviations methods.
\end{abstract}

\maketitle

\section{Introduction}
\label{sec:intro}

A crucial endeavour of statistical mechanics concerns the description of the
response of a system to an external force perturbing the baseline dynamics, in terms
of physically accessible quantities. The wealth of results collected in this
long-standing research field has been organized within the general framework of
response theory. Namely, response theory represents a variety of technical tools enabling one to
compute the change, when the perturbation is added, in the expectation value of an
observable from the knowledge of the invariant measure of the unperturbed dynamics and of the structure of the
perturbation. One first attempt in this direction dates back to the seminal work of
Kubo \cite{Kubo2}, who introduced a perturbative theory tackling the derivation of
response formulae for Hamiltonian thermostatted systems subjected to
an external force.
Kubo's derivation unveiled, in particular, one version of the celebrated
Fluctuation-Dissipation Theorem (FDT) \cite{Kubo, BPRV}, which establishes a
conceptually rich and a practically useful connection between the (linear) response
of a system to external perturbations and the equilibrium fluctuations of suitably
defined observables \cite{Zub2}.
As discussed in Ref. \cite{Falc}, the chance of bridging the external forcing with
the fluctuations computed along the unperturbed dynamics mostly relies on the assumption
that the invariant measure of the unperturbed system is smooth. Therefore, when
considering chaotic dissipative dynamical systems, equipped with an invariant
measure with support on a strange attractor, the FDT no longer holds, in general, as
shown by Ruelle in Ref. \cite{R3}. To first order in the perturbation, in fact, the response formula
results, namely, from the sum of two terms. The first one can be cast into a correlation
function evaluated with respect to the unperturbed measure along unstable and
neutral manifolds, and can be regarded as the natural nonequilibrium extension of the
correlation function occurring in equilibrium theory. The second
term, instead, involves the computation of the statistical properties of the dynamics along
the stable manifold: this has no counterpart in equilibrium.
The foreseen breaking of the FDT for dissipative deterministic dynamics rules out,
in principle, the possibility of interpreting the response solely from the knowledge
of the unperturbed steady state dynamics, thus hindering the investigation of such systems which are ubiquitous
in nature. Note that, just to take a relevant example in a geophysical setting, the
validity of a FDT would allow, in principle, to compute forced climate change from
the investigation of natural climate variability. Nonetheless, few examples have
been reported in the literature, witnessing that the FDT seemingly survive also
beyond equilibrium \cite{Grits,Grits2}. \\
The problem of lack of smoothness of the invariant measure can be circumvented through different routes. One may, in first instance, introduce some noise on top
of the deterministic dissipative dynamics, so as to mimic, say, the effect of
round-off errors in numerical algorithms or the presence of unresolved scales
\cite{Abra}. The latter perspective is adopted, in particular, in the approach pursued by Zwanzig
\cite{Zwanzig} in his projection operator formalism, in which the dynamics of the
macroscopic, physically relevant, variables is triggered also by a term, typically
regarded as noise, which echoes the intrinsic coupling with the neglected, more
microscopic, degrees of freedom.
Hence, the introduction of a small amount of noise, which can be motivated on physical grounds, allows one to restore the invoked smoothness of the invariant
measure, thus making the FDT still applicable \cite{Wou,Wou2}. \\
Nevertheless, even without adding noise, some recent findings on the FDT
for dissipative dynamical systems \cite{CRV} confirm that the link
between response and fluctuations computed with respect to the unperturbed invariant measure can
be safely restored for most of systems of interest in Physics.
The reason is that, in statistical mechanics, one typically deals with projected dynamics, and these are
associated with regular probability distributions in the
corresponding lower dimensional spaces.
A major focus of this paper is on the derivation of a general formalism able to
encompass former results obtained for deterministic as well as stochastic dynamics. We aim, in
particular, at providing a more straightforward interpretation of the response
formalism in terms of observables of clearer physical relevance. In this direction,
in a recent work \cite{CMW} focusing on the stochastic dynamics of open mesoscopic
systems, it was shown that the response of a generic observable can be cast, at the
various orders, in terms of correlation functions featuring two main time-extensive
quantities: the (excess) entropy flux, which can be understood within the standard thermodynamic framework, and a novel term, called dynamical activity,
which still lacks a conclusive physical interpretation.
This work represents a step forward along that direction: we take the point of view
of deterministic mechanics and provide some of the terms appearing in
Ruelle's original response formula with a physical content. To this aim, as
discussed above, we assume the existence of a smooth invariant measure, and discuss
the resulting response formula up to the second order. We thus show that, at the
first order, a prominent role is played by the entropy production, hence recovering former results obtained for deterministic \cite{Evans1} as well as stochastic dynamics \cite{MN}.
At the second order, the expectation value of a given observable is ruled by two time-symmetric terms. The first term is related to the first correction to the average entropy production of the system \cite{R1}, whereas the second term keeps track of the phase space fluctuations of the entropy production. Although an obvious correspondence with time-symmetric dynamical activity, introduced in \cite{CMW} to deal with stochastic diffusive dynamics, is still missing, the formalism developed below allows one to pave the a promising bridge between different methods used in response theory. \\
The work is organized as follows.\\
In Section \ref{sec:sec1} we present a rewriting of the linear response introduced in Ref. \cite{R1},
and we evidence the role the so-called dissipative flux as being the relevant observable entering the linear order response formalism.\\
In Section \ref{sec:sec2} we show how our results can be used to ascertain and
interpret the linear response to perturbations in various deterministic settings. \\
The case of stochastic dynamics is analyzed in Section \ref{sec:sec3}, in which we show that our framework is consistent with the response formulae obtained from large deviations methods. \\
In Section \ref{sec:sec4} we present some results pertaining to the second order terms and shed light on the onset of two different quantities. \\
Conclusions are drawn in Sec. \ref{sec:concl}.

\section{Entropy production and Linear Response Theory}
\label{sec:sec1}

We start by concisely recalling Ruelle's approach to response theory in
deterministic dynamical systems \cite{R3,R1,R2}.
Let ($\mathcal{U},S_o^t,\mu_o$) be a dynamical system, with $\mathcal{U}$ denoting a
compact phase space, $S_o^t:\mathcal{U}\rightarrow \mathcal{U}$ a one-parameter
group of diffeomorphisms and $\mu_o$ the invariant natural measure. Let also $x \in \mathcal{U}$ denote a generic phase space point
$x=(q_1,...,q_N,p_1,...,p_N)$.
We assume, from here onwards, that the measure $\mu_o$ is absolutely continuous with
respect to the Lebesgue measure, i.e. it is equipped with an invariant density
$\rho_o(x)$ such that
\be
\mu_o(dx)=\rho_o(x) dx \quad .
\ee
We consider, then, the effect of adding a small (possibly time-depedent) perturbation $f_t(x)$ at time $t = 0$, which induces the following structure of the equations of dynamics:
\be
\dot{x_t}=F(x_t)+ f_t(x_t) \label{dyn} \quad ,
\ee
where we used the shorthand notation $x_t=x(t)$, and $F$ denotes the drift
of the unperturbed dynamics. In the following, when convenient, we will also
split the perturbation as:
$f_t(x)=h_t X(x) $, where $h_t$ is a time modulation of the phase space function
$X(x)$. Because the perturbation is small, one may follow a
perturbative approach to express the change in measure $\mu_o$ induced by the
perturbation itself.
Following Ruelle's arguments \cite{R1,LC}, the response in a generic observable
$B:\mathcal{U}\rightarrow\mathbb{R}$ may be expressed as a perturbative expansion:
$$
\lan B(t) \ran^h=\lan B(t)\ran^o+\sum_{n=1}^\infty \lan\delta B (t) \ran_n^h
$$
where the superscript ``h'' on the lhs of the above equality is meant to recall that the average is
computed wrt the perturbed density, whereas, on the rhs, the term $\lan
B(t)\ran^o=\int dx_0 \rho_o(x_0) B(x_t)$ denotes the expectation value of $B(t)$ wrt the unperturbed density. The terms $\lan\delta B (t) \ran_n^h$
attain the formal structure:
\be
\lan\delta B (t) \ran_n^h= \int_{-\infty}^{+\infty} ds_n...\int_{-\infty}^{+\infty}
ds_1 G^{(n)}(s_1,...,s_n) h_{t-s_1}...h_{t-s_n} \label{general} \quad .
\ee
The $n^{th}$ order Green function, in particular, can be read off explicitly:
\bea
G^{(n)}(s_1,...,s_n)&=& \int dx \rho_o(x) \theta(s_1)...\theta(s_n-s_{n-1}) \times \nonumber\\
&\times& \Lambda\Pi(s_n-s_{n-1})...\Lambda\Pi(s_2-s_{1})\Lambda\Pi(s_1) B(x) \label{green}
\quad ,
\eea
where $\theta(t)$ denotes the heaviside step function, and with:
\be
\Lambda \Phi=X(x)\cdot\frac{\partial}{\partial x}\Phi \hskip 30pt \mbox{and} \hskip 30pt
\Pi(t)\Phi=\Phi \circ S_o^t \label{pi} \quad.
\ee
At the linear order, Eq. (\ref{general}) simplifies into:
\be
\lan\delta B (t) \ran_1^h = \int_{0}^{+\infty}  G^{(1)}(t-s) h_{s} ds=\int_{0}^{t} R(t-s) h_s ds \label{gen1st}
\ee
where $R(t)$ denotes the so-called \textit{response function} and $G^{(1)}(t)=\theta(t)R(t)$.
We recall that, by using the Kramers-Kronig relations \cite{Lucar}, the principle of
causality boils down, in the frequency domain, to the following alternative
relations between the Fourier transforms $\chi(\nu)=\mathcal{F}(G(t))$ and
$\hat{R}(\nu)=\mathcal{F}(R(t))$:
\be
\hat{R}(\nu)=2 \mathrm{Re}\{\chi(\nu)\} \quad \mathrm{or} \quad
\hat{R}(\nu)=2 \mathrm{Im}\{\chi(\nu)\} \quad , \nonumber
\ee
depending on whether the response function is, respectively, even or odd under the
time-reversal, cf. \cite{Kubo,LC}.
A simple calculation shows, then, that, at the first order, the expansion
(\ref{general}) leads to the familiar FDT:
\bea
\lan\delta B (t) \ran_1^h &=&\int_{0}^t h_s ds \int \rho_o(x_0)
X(x_0)\left(\frac{\partial}{\partial x_0} B(x_{t-s})\right) dx_0 =\nonumber\\
&=& \int_{0}^t  ds \int \sigma_s(x_s) B(x_t) \rho_o(x_0) dx_0= \lan B(x_t) S(\omega)
\ran^o\label{1} \quad ,
\eea
where we introduced the \textit{dissipative flux} $\sigma_s(x_s)$:
\be
\sigma_s(x_s)=h_s\gamma(x_s) \label{sigma0} \quad,
\ee
with
\be
\gamma(x)=-\frac{1}{\rho_o(x)}\left[\frac{\partial}{\partial x}\cdot(X(x)
\rho_o(x))\right] \label{agarw} \quad .
\ee
In Eq. (\ref{1}) we have denoted by $S(\omega)$ the integral of $\sigma_s$ over the path $\omega = (x_s,
s\in [0,t])$ started from $x_0$: $S(\omega)$ is a phase function which, under
suitable assumptions (e.g., when the perturbation is conservative, cf. Sec.
\ref{sec:sec2}), can be associated to the total entropy produced along
the path. Moreover, from Eqs. (\ref{gen1st}) and (\ref{1}), one also obtains the formal
expression for the response function:
\be
R(t-s)= \lan B(x_t) \gamma(x_s) \ran^o \label{R} \quad .
\ee
The phase function $\sigma_s(x_s)$, introduced in (\ref{sigma0}), can be split, using Eq.
(\ref{agarw}), as follows:
\be
\sigma_s(x_s)= \underbrace{-\frac{\partial }{\partial x_s}\cdot
f_s(x_s)}_{\sigma_A}+\underbrace{f_s(x_s)\cdot \frac{\partial}{\partial x_s}
\left(-\log \rho_o(x_s)\right)}_{\sigma_B} \label{sigmadef} \quad .
\ee
While the term $\sigma_A$, in Eq. (\ref{sigmadef}), corresponds to a purely
dissipative contribution, induced by nonconservative perturbations leading to
dissipative dynamics, the term $\sigma_B$, in turn, which contains the function $(-\log
\rho_o)$ (referred to, in the literature, as the \textit{information
potential}, cf. Sec. \ref{sec:sec3.2}), is related to the total entropy produced by the perturbation and released into the environment, regardless of whether the perturbation is conservative of not (this is, actually,
the term originally introduced in Kubo's theory \cite{Kubo}).
Noticeably, the expression in Eq. (\ref{sigmadef}) recovers the original Agarwal
formula \cite{Agarw}, which hence corresponds to the
linear order contribution of Ruelle's formal expansion. As also outlined in
\cite{Baies}, the use of Eq. (\ref{sigmadef}) is typically hindered by the lack of
the knowledge of the reference density $\rho_o$ and, possibly, of the details of the
perturbed dynamics triggered by the phase function $X(x)$.
Nevertheless, the Agarwal formula (\ref{sigmadef}) has enjoyed a growing popularity
in the literature and was, since then, derived following different routes, cf.
\cite{BPRV,Grits, Weid, Boch1,Boch2,Hang}. \\
Moreover, we also observe that the observable $S(\omega)$ formally resembles
the structure of the so-called \textit{Dissipation Function} introduced in Ref.
\cite{Evans1} (cf. Eq. (9) therein), and for which transient and steady state Fluctuation Relations have
been proven under rather general conditions \cite{Evans2}.\\
Thus, Eq. (\ref{sigmadef}) suggests that, to first order, the linear
response can be cast as an equilibrium time correlation function between the chosen
observable and $\sigma_s(x_s)$, which includes the two different aforementioned
source terms $\sigma_A$ and $\sigma_B$. For weakly perturbed dynamical systems, the
proposal of expressing the response function as a correlation between the chosen
observable and the dissipative flux $\sigma_s$ can be already traced back to the
seminal works of McLennan \cite{McL} and Zubarev\cite{Zub}. We remark, moreover, that this
general framework is not restricted to deviations from an \textit{equilibrium}
reference state only, but holds for any steady state equipped with a smooth
invariant density $\rho_o(x)$. In particular, as it will be discussed in Sec.
\ref{sec:sec2}, our approach allows one to make a bridge with the results outlined in Ref. \cite{CRV}, where the authors considered a reference invariant SRB measure equipped with a smooth marginal probability density, resulting from the projection of the full
SRB measure along the direction of the initial impulsive perturbation.
We also point out that the response formalism introduced above, based on the use of the operators (\ref{pi}), is prone to be also used in the set-up of stochastic dynamics \cite{R1}: the application of the method to diffusive systems is deferred to Sec. \ref{sec:sec3}.\\
It is worth mentioning two basic properties characterizing  the dissipative flux $\sigma_s(x_s)$.\\
First, note that the expectation $\langle \sigma_s(x_s)\rangle^o$, computed with the unperturbed density, vanishes:
\be
\int \rho_o(x) \sigma_s(x) dx = - \int dx \left[\frac{\partial}{\partial
x}\cdot(X(x) \rho_o(x))\right] =0 \quad . \label{equil}
\ee
The relation (\ref{equil}) allows one, hence, to interpret the observable $\sigma_s(x_s)$ as an ``excess'' dissipative flux of the perturbed process with respect to the unperturbed one. In more physical terms, $\sigma_s(x_s)$ can be thus regarded as the surplus of entropy production, due to the perturbation, with respect to the ``housekeeping'' heat flux needed to maintain the steady state \cite{Maes1}.
Notice that to ensure that the response function be integrable, one typically requires the correlations to decay sufficiently fast \cite{R1}.\\
Next, let us consider the time-reversal symmetry property of $\sigma_s(x_s)$.
To this aim, following Roberts \textit{et al.} \cite{RQ}, we define the involution
$G$ as:
\be
D_G \cdot f_t = -f_t \circ G \quad , \label{G}
\ee
with $G\circ G= 1$, where $D_G$ denotes the Jacobian matrix of $G$
\footnote{It is worth recalling that, in the discrete time case, the dynamical flow is replaced by a mapping $M:\mathcal{U}\rightarrow\mathcal{U}$, and Eq. (\ref{G}), correspondingly, takes the form: $G\circ M\circ G=M^{-1}$ \cite{CK,CR}.}.\\
A dynamical system is said to be \textit{reversible} if there exists an involution $G:\mathcal{U}\rightarrow\mathcal{U}$ fulfilling Eq. (\ref{G}) (i.e. it reverses the direction of time).
Using (\ref{G}) one thus finds:
\be
\sigma(Gx)=-\sigma(x) \label{odd} \quad ,
\ee
where we assumed the unperturbed density to be invariant under the
involution $G$, $\rho_o(x)=\rho_o(Gx)$ \footnote{The property of invariance of $\rho_o$ under the involution $G$ was called, in Ref.
\cite{CR}, as \textit{phase space detailed balance}, because it reduces, via a suitable projection onto the space of stochastic dynamics, to the \textit{detailed balance} relation, which stands as the hallmark of an equilibrium stochastic dynamics \cite{CMW}.}.
Equation (\ref{odd}) indicates that the phase function $\sigma_t(x)$ is, as expected, odd under time-reversal, and so is, therefore, its time integrated value $S(\omega)$, over the path $\omega$ \cite{CK}:
$$
S(G\omega)=-S(\omega)
$$

\section{Deterministic dynamics}
\label{sec:sec2}

The formalism developed in Sec. \ref{sec:sec1} can be applied to a wealth of
different physical situations, in which one considers the effect of adding a small
perturbation to a reference dynamics, enjoying a steady state. It is worth remarking
that, in order to apply the formalism introduced above, one merely requires that the
reference steady state be equipped with a density.
In this Section we will investigate the application of the formalism described above
in deterministic dynamical systems, enjoying either an equilibrium or a
nonequilibrium steady states. We will restrict ourselves to the linear case.

\subsection{Hamiltonian dynamics}

It is instructive to address the case in which the reference microscopic dynamics is
Hamiltonian and the steady state is an equilibrium one.
Thus, we consider the effect of adding a small conservative perturbation, to be
expressed as the gradient of a scalar potential function $V(x)$. This corresponds to
replacing the Hamiltonian $H_0(x)$ as:
$$
H_0(x) \rightarrow H_0(x)-h_t V(x) \quad .
$$
Therefore, in Eq. (\ref{dyn}), one sets $F(x)=S \nabla H_0(x)$, where $S$ denotes
the symplectic matrix, and $f_t(x)= - h_t S \nabla V(x)$.
Next, by taking, for simplicity, the canonical distribution as the reference equilibrium density, i.e. $\rho_o(x)=Z^{-1} e^{-\beta H_0(x)}$ (where $Z$ denotes the canonical partition function and $\beta=1/k_B T$ is the inverse temperature), a straightforward calculation
reveals that the term $\sigma_A$ in (\ref{sigmadef}) vanishes and the term $\sigma_B$
attains the form:
\be
\sigma_s(x_s)=\beta h_s \dot{V}(x_s) \label{sigma}  \quad ,
\ee
where we introduced the shorthand notation $\dot{B}(x_t)=d B(x_t)/dt$.
Thus, by inserting (\ref{sigma}) in (\ref{1}), one finds:
\bea
\lan\delta B (t) \ran_1^h&=& \beta \int \rho(x_0) dx_0 B(x_t) \int_{0}^t ds
\dot{V}(x_s) h_s \nonumber \\
&=&\beta \int \rho(x_0) dx_0 B(x_t)\left[\left(V(x_t)h_t-V(x_0)h_0\right)-
\int_{0}^t ds V(x_s)\dot{h}_s\right] \nonumber\\
&=& \beta \lan S(\omega) B(x_t) \ran^o \label{Spart}  \quad .
\eea
Therefore, for Hamiltonian dynamics, Eq. (\ref{R}) reduces to the classical Green-Kubo form
\cite{Kubo,BPRV,Maes1}:
\be
R(t-s)=\beta \frac{d}{ds}\lan B(x_t) V(x_s) \ran^o \label{Rham} \quad .
\ee
In (\ref{Spart}), the term $\left(V(x_t)h_t-V(x_0)h_0\right)$ corresponds to the
extra change of energy in the environment due to the perturbation, whereas
$\int_{0}^t ds V(s)\dot{h}_s$ is the work done by the perturbation. Therefore, the
case of conservative perturbations shows that the linear response term can be
effectively cast into an equilibrium correlation between the given observable and
the total entropy produced, over the path, by the perturbation and released into
the environment, cf. \cite{Baies, Maes1}.
Let us consider a reference Hamiltonian given by $H_0=\sum_{i=1}^N p_i^2/(2m)$ (with $m$ denoting the mass of the particles), which, after the addiction of a time independent, homogeneous, electric field $E$, takes the form $H_0 \rightarrow H_0-\sum_{i=1}^N \kappa_i (q_i \cdot E)$, with $\kappa_i$ denoting the charge of the $i$-th particle. The application of Eq. (\ref{agarw}) leads, thus, to the standard Kubo expression $\sigma(x)=\beta E \cdot J$, where the electric current $J$ takes the form $J=\sum_{i=1}^N (p_i/m\cdot \kappa_i)$.

\subsection{Dissipative dynamics}

Let us consider, next, the case of a dissipative steady state dynamics, perturbed by a small impulsive perturbation, which modifies the
initial condition as $x_0 \rightarrow x_0 +\delta x_0$.
Therefore, by setting $f_s=\delta(s) \delta x_0$, one finds:
\be
\sigma_s(x_s)=-\delta(s) \left(\frac{\partial \log \rho(x_s)}{\partial x_s}\cdot
\delta x_0\right) \label{diss} \quad ,
\ee
which leads to the response formula outlined in Ref. \cite{CRV,BLV}.
One natural objection to the expression (\ref{diss}) points to the fact that the invariant
measure of a chaotic dissipative system $\mu_o$ is singular with respect
to the Lebesgue measure, and is typically supported on a fractal attractor. This may,
hence, prevent the application of the approach outlined in Sec. \ref{sec:sec1} to
dissipative systems. Indeed, the standard FDT ensures
that the statistical features of a perturbation are related to the statistical
properties of the unperturbed
system, but that cannot be the case, in general, in dissipative systems. The reason
is that,
given an initial state $x_0$ on the attractor and a generic impulsive
perturbation $\delta x_0$, the perturbed
initial state $\tilde x_0 = x_0 + \delta x_0$ and its time evolution may lie outside the
support of
the measure, hence their statistical properties cannot be expressed by $\mu$, which
attributes vanishing
probability to such states. Nevertheless, in Ref. \cite{CRV} it was shown that a
different route is possible to compute the response, which is worth recalling
shortly here.
By denoting as $\rho_o(x_0;\delta x_0)=\rho_o(x_0-\delta x_0)$ the perturbed initial density, and by $W(x_0,0\rightarrow x_t,t)$ the transition probability determined by the dynamics, one may express the response of the coordinate $x^i$ as:
\be
\lan\delta x^i (t)\ran^h = \int \int  x^i_t
\left[ \rho_o(x_0-\delta x_0)-\rho_o(x_0) \right]
W(x_0,0\rightarrow x_t,t) dx_0 dx_t \quad .
\label{V0}
\ee
By also assuming, for sake of simplicity, that all components of $\delta x_0$ vanish, except the $i$-th component, one finds that Eq.
(\ref{V0}) can be written as:
\bea
\lan\delta x^i(t)\ran^h  = \int x^i_t (\widetilde{\rho_t}(x^i;\delta
x_0)-\widetilde{\rho}_o(x^i_t)) dx^i_t \label{V2} \quad ,
\eea
where $\widetilde{\rho}_o(x^i_t)$ and $\widetilde{\rho_t}$ are the marginal
probability
distributions defined by:
$$
\widetilde{\rho}_o(x^i_t)=\int \rho_o(x_t) \prod_{j\neq i}dx^j_t \quad ,
\qquad
\widetilde{\rho_t}(x^i_t;\delta x_0)=\int \rho_t(x_0;\delta x_0) \prod_{j\neq
i}dx^j_t \quad.
$$
Thus, since projected singular measures are expected to be smooth \cite{CRV,Leb}, especially
if the dimension of
the projected space is sensibly smaller than that of the original space, one finds that a FDT, written in the form of Eq. (\ref{1}), can typically be extended to a large fraction of the dissipative deterministic systems of interest in Physics.

\section{Stochastic diffusions and large deviations}
\label{sec:sec3}

Let us now turn our attention to stochastic dynamics. In general, the presence of noise allows
one to characterize the steady state dynamics, even in presence of dissipation, by
regular probability densities, thus overcoming the typical difficulties encountered
in deterministic dynamical systems. A detailed analysis of the response formulae valid for Markovian Langevin-type stochastic differential equations is given in Ref. \cite{R4}, where an expansion formally resembling the structure of Eqs. (\ref{green}) and (\ref{pi}) is developed. In particular, in Ref. \cite{R4}, Ruelle shows that, in the zero noise limit, the various terms of the expansion reproduce the corresponding order terms pertaining to the deterministic dynamics discussed in Sec. \ref{sec:sec1}. This occurs because, under suitable assumptions, the SRB states are stable under small random perturbations \cite{Young, Kifer}. It is therefore tempting to use the formalism of Sec. \ref{sec:sec1} to describe stochastic models amenable to an analytical solution, so as to compare our results with those obtained using other methods, e.g. the path-integral formalism described in Refs. \cite{CMW, Baies}.
We will thus focus, first, on stochastic diffusion processes described by
overdamped Langevin equations, in which one disregards inertial effects, thus letting
forces be proportional to velocities rather than to accelerations
\cite{Maes1,MNW}. These processes correspond to the high damping limits of the underdamped, or intertial, stochastic dynamics, whose analysis is deferred to Sec. \ref{sec:sec3.3}.
Let us start considering overdamped diffusion processes for the state $x\in {\mathbb R}^n$, defined in the It\^o sense by:
\be
\dot{x}_t = \chi \cdot [F(x_t)+ f_t(x_t)] + \nabla \cdot D(x_t)+\sqrt{2
D(x_t)}\,\xi_t \quad , \label{diff}
\ee
where $\xi_t$ denotes a standard white noise and $f_t$ denotes the perturbation to the reference dynamics.
The mobility $\chi$ and the diffusion constant $D$ are strictly positive (symmetric)
$n\times n$-matrices, which, provided that the system is in contact with a
thermostat at inverse temperature $\beta>0$, are connected via the well-known
Einstein relation $\chi=\beta D$.
The force $F$ denotes the drift of the reference, unperturbed dynamics, and can be expressed as:
\be
F = F_{nc}- \nabla U  \label{force} \quad ,
\ee
where $F_{nc}$ denotes a nonconservative force pulling the reference dynamics out of equilibrium, while $U$ is the energy of the system.
The Fokker-Planck equation for the time dependent density $\rho_t$, relative to the diffusion process described by (\ref{diff}), reads
\be
 \frac{\partial \rho_t}{\partial t}(x_t) = -\nabla \cdot j_{\rho} \hskip 2pt, \hskip 8pt
\mbox{with} \hskip 8pt j_{\rho}=[\chi (F+ f_t)\rho_t(x_t)-\frac{\chi}{\beta}\nabla\rho_t(x_t)] \label{fp}   \quad ,
\ee
where $j_{\rho}$ denotes the probability current \cite{risken}.
Rather than attempting a direct analytical solution of Eq. (\ref{fp}), one may tackle the analysis of Eq. (\ref{diff}) from the rather different standpoint of large deviations theory \cite{CMW,dembo}. We shorty recapitulate the main steps of the
derivation, cf. Refs. \cite{CMW,MN} and references therein for a more detailed
discussion.
The key idea is to determine the perturbed probability density through its embedding
in the path-space distribution. That is, given the (random) paths $\omega = (x(s),
s\in [0,t])$, one may connect the distribution $P$ on paths starting from $\rho_o$
and subjected to the perturbation $f_t$, with the reference distribution $P^o$
concerning paths starting from $\rho_o$ and undergoing the reference dynamics, via
the formula:
\begin{equation}\label{act}
P(\omega) =e^{-\mathcal{A}(\omega)}\,P^o(\omega) \quad .
\end{equation}
The relation (\ref{act}) defines the action $\mathcal{A}(\omega)$, which is typically local in
space-time and is, thus, similar to Hamiltonians or Lagrangians encountered in
equilibrium statistical mechanics, see e.g. \cite{poincare}.
One can also verify that
\[
\mathcal{A} = (\mathcal{T} - S)/2 \quad,
\]
where $\mathcal{T}(\omega)$ and $S(\omega)$ are
path-dependent quantities corresponding, respectively, to the time-symmetric and
time-antisymmetric components of the action. That is, defining the
time-reversal operator $g$ as:
\be
 g \omega = ((\pi x)_{t-s}, 0\leq s\leq t) \quad , \label{invstoc}
\ee
(with $\pi{x}$ equal to $x$ except for flipping any other variable with negative
parity under time-reversal), one can write:
\be
 S(\omega) = \mathcal{A}(g\omega) - \mathcal{A}(\omega) \hskip 5pt, \quad \mathcal{T}(\omega) = \mathcal{A}(g\omega) + \mathcal{A}(\omega) \label{revers} \quad .
\ee
The quantity $S(\omega)$, under the assumption of local detailed balance \cite{Katz}, is the entropy flux triggered by the perturbation and released into the environment \cite{CMW}. On the other hand, the quantity $\mathcal{T}(\omega)$ is referred to in the literature as \textit{the dynamical activity} or \textit{traffic} \cite{Maes1,MNW} and appears to be much more concerned with kinetics than it is embedded into
thermodynamics. For example, in the set-up of Markov jump processes, $\mathcal{T}(\omega)$ is suitable to a physical interpretation: it measures how the escape rate from a trajectory $\omega$ changes when the perturbation $f_t$ is added \cite{Maes1}.\\
A simple calculation unveils the following general expression for the action pertaining to the process described by Eq. (\ref{diff}):
\be
\hskip -20pt \mathcal{A}(\omega)=\frac{\beta}{2}\int_0^t ds \left[f_s \cdot \chi F + \nabla \cdot (D f_s)+
\frac{1}{2} f_s \cdot \chi f_s\right] - \frac{\beta}{2}\int_0^t \id x_s \circ f_s
\label{A}
\ee
where the last stochastic integral (with the $\circ$) is in the sense of
Stratonovich. From (\ref{revers}) and (\ref{A}), one can derive the following
expressions for $S(\omega)$ and $\mathcal{T}(\omega)$:
\be
S(\omega) = \beta \int_0^t \id x_s \circ f_s \hskip 17pt \mbox{and} \hskip 17pt \mathcal{T}(\omega) = \mathcal{T}_1+\mathcal{T}_2  \nonumber \quad ,
\ee
with
\be
\mathcal{T}_1 =  \beta \int_0^t ds \left[f_s \cdot \chi F + \nabla \cdot (D
f_s)\right] \hskip 17pt \mbox{and} \hskip 17pt \mathcal{T}_2 =  \frac{\beta}{2} \int_0^t ds  f_s \cdot \chi f_s \nonumber \quad .
\ee
If the chosen observable $B$ is endowed with an even kinematical parity, the following linear response formula can be thus established \cite{Maes1}:
\be
\hskip -50pt \lan\delta B (t)\ran_1^h= \lan B(x_t) S(\omega)\ran^{o}=-\lan B(x_0) S(\omega)\ran^{o} = -\int dx_0 \rho_o(x_0)
 B(x_0) \lan S(\omega) \ran^o_{x_0}\label{1LG} \hskip 5pt .
\ee
The expression (\ref{1LG}), thus, inherits the structure of the response formula (\ref{1}) obtained for deterministic systems.
The quantity $\lan S \ran^o_{x_0}$, in Eq. (\ref{1LG}), denotes the conditional expectation of the entropy flux $S(\omega)$ over $[0,t]$ given that the path started from the state $x_0$.
It can also be written as \cite{CMW,MN}::
\be
\lan S\ran^o_{x_0}= \beta \int_0^t \langle w(x_s) \rangle^o_{x_0} ds \label{w} \quad ,
\ee
where $w(x_s)$ corresponds to the instantaneous (time-antisymmetric, random)
work made by the perturbation $f_t$.
The analysis of some specific models and examples comes next.

\subsection{Expansion around detailed balance dynamics}
\label{sec:sec3.1}

An interesting example is obtained by considering overdamped diffusion processes whose reference dynamics is a (equilibrium) detailed balance dynamics \cite{CMW}, i.e. $F_{nc}=0$ in Eq. (\ref{force}). Let us also take
the reference distribution of states to be the equilibrium one, $\rho_o(x) \propto
e^{-\beta U(x)}$.\\
To further simplify the problem, let us also assume the perturbation $f$ to be time
independent and the matrices $\chi$ and $D$ to be independent of $x$.
The quantity $w(x_s)$ pertaining to this stochastic dynamics can be explicitly computed \cite{CMW,MN}:
\be
 w(x) =\frac{\chi}{\beta}\nabla \cdot f - \chi f\cdot \nabla U
\label{weq}
\ee
By inserting the expression (\ref{weq}) into Eq. (\ref{w}) and by further using Eq. (\ref{1LG}), one readily obtains the FDT for the process under consideration.
Alternatively, one may adopt the formalism detailed in Sec. (\ref{sec:sec1}) to derive the linear response. To this aim, by computing
$\sigma_s(x_s)$, given in Eqs. (\ref{sigma0}) and (\ref{agarw}), one immediately recovers
the expression for $\langle w(x_s) \rangle^o_{x_0}$ in (\ref{w}) hence recovering the above response formula.\\
It is also instructive to consider the case of a (conservative) perturbation
changing the potential $U$ into $U-h_t V$.
For the case under consideration, the general response formula holds \cite{Maes1}:
\be
\hskip -40pt R(t-s)=\frac{\beta}{2}\frac{d}{ds}\lan V(x_s)B(x_t)\ran^o-\frac{\beta}{2}\lan
LV(x_s) B(x_t)\ran^o=-\beta\lan LV(x_s) B(x_t)\ran^o \hskip 10pt , \label{genresp}
\ee
where we introduced the (backward) generator $L$ of the process\footnote{Hence, from the perspective of large deviations theory, the equilibrium FDT may also be equivalently cast into a \textit{fluctuation-activity} relation, by exploiting the properties of the only time-symmetric term of the action.}, defined as
$$
L=-\chi \nabla U\cdot \nabla+\frac{\chi}{\beta}\nabla^2 \quad .
$$
To derive the last equality in Eq. (\ref{genresp}), one uses the time-reversal symmetry of the equilibrium correlations and the properties of the adjoint generator $L^*=L$: \footnote{$L^*$ is defined with the help of the stationary distribution $\rho_o$: for any two state functions $f$ and $g$, $L^*$ is such that $\int dx \rho_o(x)g(x)L^* f(x)=\int dx \rho_o(x)f(x)L g(x)$. For detailed balance dynamics, in particular, one has $L^*=\pi L \pi$, where $\pi$ flips the variables which are odd under time-reversal, cf. Eq. (\ref{invstoc}).}
\be
\hskip -54pt \lan L^*V(x_s) B(x_t)\ran^o= \lan V(x_s) L B(x_t)\ran^o= \frac{d}{dt} \lan V(x_s) B(x_t)\ran^o =-\frac{d}{ds} \lan V(x_s) B(x_t)\ran^o \label{rel} \hskip 5pt,
\ee
for $s<t$.
On the other hand, when turning back to the formalism of Sec. \ref{sec:sec1},
it is not difficult to obtain the following expression for the quantity $\gamma(x)$
defined in (\ref{agarw}):
\be
\gamma(x)=\beta\chi\nabla U\cdot\nabla V-\chi\nabla^2 V\label{Leq} \quad .
\ee
Thus, by plugging Eq. (\ref{Leq}) into Eq. (\ref{R}), one immediately recovers Eq. (\ref{genresp}).\\
The chosen examples, pertaining to equilibrium dynamics subjected to time-independent perturbations, corroborate, hence, the equivalence between the linear response formulae derived either from the deterministic Ruelle's expansion or from the above introduced large deviation formalism.

\subsection{Nonequilibrium steady states}
\label{sec:sec3.2}

By setting $F_{nc}\neq 0$, in Eq. (\ref{diff}), one spoils the time-reversibility of the reference dynamics. Therefore, given enough time, the reference
dynamics reaches a nonequilibrium steady state described by an invariant density
$\rho_o(x)$. The latter is typically not known, nevertheless the approach traced
in Sec. \ref{sec:sec2} allows one to obtain linear response formulae recovering the corresponding expressions obtained via the path-integral formulation outlined above.
In the steady state, one can use the definition of the probability current given in Eq.
(\ref{fp}), to define the \textit{information potential} $\mathcal{I}_{\rho_o}$ \cite{Baies,Prost} as:
\be
\mathcal{I}_{\rho_o}=-\frac{d \log \rho_o}{dx}=\frac{\beta}{\chi}u-\beta F \quad,
\label{infpot}
\ee
where $u\equiv j_{\rho_o}/\rho_o$ denotes a probability velocity.
Therefore, using Eqs. (\ref{R}) and (\ref{sigmadef}), with
Eq. (\ref{infpot}), one finds the general response function for nonequilibrium overdamped
diffusion processes:
\be
\hskip -30pt R(t-s)=\chi\lan \left[-\frac{d}{dx_s} \cdot f(x_s)+
\mathcal{I}_{\rho_o}(x_s)\cdot f(x_s)\right] B(x_t)\ran^o \quad .
\ee
In particular, if the perturbation takes the gradient form $f=\nabla V$, an easy calculation yields:
\be
R(t-s)=\beta\lan \left(u(x_s)\cdot\nabla V(x_s)\right)B(x_t)\ran^o-\beta \lan LV(x_s) B(x_t)\ran^o \quad , \label{LV}
\ee
with $L=\chi F\cdot \nabla+\chi/\beta\nabla^2$.
Next, by using the relations (\ref{rel}), with
$$
L^*=-\chi F\cdot \nabla+\frac{\chi}{\beta}\nabla^2+2\frac{\chi}{\beta}\nabla (\log \rho_o)\cdot \nabla=L-2u\cdot \nabla \quad ,
$$
one can suitably transform Eq. (\ref{LV}) into the equivalent form:
\be
R(t-s)=-\beta\lan \left(u(x_s)\cdot\nabla V(x_s)\right)B(x_t)\ran^o+\beta \frac{d}{ds}\lan  B(x_t) V(x_s)\ran^o \quad , \label{LV2}
\ee
which successfully recovers Eq. (26) of \cite{Baies}. It is worth remarking that the
function $u(x)$, in (\ref{infpot}), is unknown in general. Nevertheless, Eq.
(\ref{LV}) is noteworthy at a formal level, for it shows that the response function
can be expressed in terms of a suitable correlation function computed wrt to
reference stationary density characterizing the nonequilibrium steady state.
One also readily notices that Eq. (\ref{LV2}) reconstructs the classical Kubo
formula (\ref{Rham}) when setting $F_{nc} = 0$ (i.e. $u=0$) or when
describing the response in a reference frame moving with drift velocity $u$, cf.
Refs. \cite{Baies,Gaw}.
Thus, the fist correlation on the rhs of Eq. (\ref{LV2}), including the function $u$, stands as the true nonequilibrium extension of the FDT to nonequilibrium steady states.

\subsection{Inertial dynamics: a solvable example}
\label{sec:sec3.3}

In this paragraph, we apply the formalism detailed in Sec. \ref{sec:sec3} to tackle the description of underdamped diffusion processes, in which
inertial effects are taken into account. We, thus, consider states
$(q,p)=(q_1,...,q_N,p_1,...,p_N) \in \mathbb{R}^{2N}$ of positions and momenta of
$N$ particles, each of which is subjected to a viscous force $(-\nu_i p_i)$, and is coupled with its own heat bath, characterized by a standard white noise $\xi_t$, with diffusion coefficient $D_i$ and
inverse temperature $\beta_i=\nu_i/D_i$, with $i=1,...,N$. We will restrict our analysis to the so-called Ornstein-Uhlenbeck diffusion process, which is amenable to an analytical solution. \\
The equations of the unperturbed dynamics read:
\bea
\dot{q}_i&=&p_i  \quad ,\nonumber\\
\dot{p}_i&=& -\frac{\partial U(q)}{\partial q_i}-\nu_i p_i+\sqrt{2 D_i}\,\xi_t \quad,
\label{underd}
\eea
with $U(q) = 1/2 m \omega_0^2 \sum_{i=1}^N q_i^2$ denoting, here, the internal energy of the particle
system. The Hamiltonian takes, hence, the structure $H_0(q,p)= \sum_{i=1}^N p_i^2/(2m)+U(q)$, and the
unperturbed (equilibrium) dynamics enjoys an invariant density of the form \cite{risken}:
\be
\rho_o(q,p)=\frac{1}{\mathcal{Z}}e^{-\beta H_0(q,p)} \quad ,\label{OU}
\ee
with $\mathcal{Z}$ denoting a normalizing factor. We consider, here again, the effect of a perturbation changing the Hamiltonian as $H_0 \rightarrow
H_0-h_t V(q)$.
By applying the formula (\ref{sigmadef}) to the set of equations (\ref{underd}),
and using Eq. (\ref{OU}), one thus obtains:
\be
\gamma(q,p)= \frac{\beta}{m} \sum_{i=1}^N \nabla_{q_i} V\cdot p_i=\frac{\beta}{m} \nabla_q V\cdot p \quad , \label{gammaOU}
\ee
where, in the last equality  of Eq. (\ref{gammaOU}), we made use of a compact notation. From Eq. (\ref{gammaOU}), one thus obtains the desired response formula:
\be
   R(t-s)=\frac{\beta}{m}\lan \left(\nabla_{q} V(q_s)\cdot p_s\right) B(q_t,p_t)
   \ran^o=\beta\frac{d}{ds}\lan B(q_t,p_t) V(q_s) \ran^o \label{under} \quad ,
\ee
which is consistent with previous derivations discussed in Refs. \cite{Baies,Maes2}.

\section{Higher order terms}\label{sec:sec4}

In this Section, we concentrate on the structure of the second order contribution to the response formulae detailed in sec. \ref{sec:sec1}. Our
aim is, again, to provide a physical interpretation of the terms appearing in the formalism beyond the standard FDT, thus complementing the formal properties studied in Ref. \cite{LC} and setting the stage for a useful bridge with the large deviation method introduced in Sec. \ref{sec:sec3}.
At the second order, Eq. (\ref{general}) attains the structure:
\be
\langle \delta B(t)\rangle_2^h =\int_{-\infty}^{\infty}ds_1\int_{-\infty}^{\infty} ds_2
G^{(2)}(s_1,s_2) h_{t-s_1} h_{t-s_2}  \quad ,
\ee
with
\be
\hskip -60pt G^{(2)}(s_1,s_2)=\theta(s_1)\theta(s_2-s_1)\int dx_0 \rho_o(x_0) X(x_0) \cdot \frac{\partial}{\partial x_0} \left[X(x_{s_2-s_1}) \cdot \frac{\partial}{\partial x_{s_2-s_1}} B(x_{s_2})\right] \label{G2}
\ee
If the reference dynamics is conservative, one finds:
\bea
\langle \delta B(t)\rangle_2^h &=& \int_{0}^{\infty}ds_1  h_{t-s_1}\int_{s_1}^{\infty}ds_2
h_{t-s_2}\times\nonumber\\
&\times&\lan\left[\gamma^{(2)}(x_0,x_{s_2-s_1}) +
\chi^{(2)}(x_0,x_{s_2-s_1})\right] B(x_{s_2})\ran^o \label{2b} \quad ,
\eea
where we introduced the second order terms $\gamma^{(2)}$ and $\chi^{(2)}$,
defined respectively as:
\bea
\gamma^{(2)}(x_0,x_s)&=& \gamma(x_0) \gamma(x_s) \label{sigma2ord} \quad ,\\
\chi^{(2)}(x_0,x_s)&=& - X(x_s) \frac{\partial \gamma(x_0)}{\partial
x_s} \quad .\label{chi2ord}
\eea
At variance with the dissipative flux $\sigma(x)$, which enters the linear response formula (\ref{1}), one notices, here, the onset of the second order quantities, $\gamma^{(2)}$ and $\chi^{(2)}$, which are even under time-reversal, as it can be easily verified by using Eq. (\ref{G}).
The time integral of the expectation value of $\gamma^{(2)}$, in Eq. (\ref{sigma2ord}), which
basically corresponds to the time correlation of the dissipative flux computed at time $t=0$ and at $t=s$, yields the leading non-vanishing contribution to the excess entropy production \cite{R1}, cf. also Eq. (\ref{equil}).
On the other hand, one also finds that the term (\ref{chi2ord}) can be rewritten as:
$$
\chi^{(2)}(x_0,x_s)=- X(x_s) \frac{\partial \gamma(x_0)}{\partial x_0}[D_{S_o^s}(x_0)]^{-1}  \quad .
$$
The $\chi^{(2)}$ term is responsible for describing the coupling between the perturbation and the gradient of the dissipative flux: its relevance stems, hence, from the presence of fluctuations of the observable $\gamma(x)$ in the phase space. In this perspective, more insight into the meaning of such term might be thus obtained by referring to the Fluctuation Theorem reported in \cite{Evans1,Evans2}.\\
From the perspective of the large deviation approach of Sec. \ref{sec:sec3}, the role of time-symmetric quantities becomes also similarly visible when going to the second order. In particular, in Ref. \cite{CMW}, it was shown that the second order term attains the structure:
\be
\langle \delta B(t)\rangle_2^h = - \frac{1}{2}\langle B(x_t)S(\omega)\mathcal{T}_1(\omega)\rangle^{o} \label{ld2ord} \quad ,
\ee
featuring the combined contribution of both the (linear order) time-symmetric and time-antisymmetric components of the action.
It is not entirely obvious to establish a neat correspondence between our second order results and those appearing in Eq. (\ref{ld2ord}), mostly because the way the expansion is performed differs between the two methods when nonlinear terms are considered. Nevertheless, it is definitely worth attempting to shed light further on the deterministic interpretation of the dynamical activity term $\mathcal{T}$, whose role in statistical mechanics has been largely unnoticed so far.

\section{Conclusions}\label{sec:concl}

The analysis of the response of statistical mechanical systems to external perturbations is of crucial relevance for
both theoretical reasons and for devising numerical and laboratory experiments. Depending on whether the underlying dynamics is Hamiltonian or
dissipative, deterministic or stochastic, a wealth of mathematical techniques have been introduced, in the literature,
to obtain perturbative response formulae. While the various mathematical formalisms settle on firm grounds, and are often prone to an algorithmic implementation, an open question addresses the physical interpretation of the terms entering the perturbation method.
In this paper we have unveiled the onset of some recurrent structures occurring in different formalisms and commented on their thermodynamic, or kinetic, foundation.\\
We have taken as starting point the response formulae proposed by D. Ruelle, through the prism of chaotic dynamical systems theory. While our approach stems from the assumption that the unperturbed dynamics is endowed with a smooth invariant measure, we have also discussed the extension to dissipative deterministic dynamics, where the smoothness of the marginal density pertaining to the projected dynamics becomes crucial. Moreover, we have highlighted the role of the dissipative flux at the first and second order of expansion, for deterministic as well as stochastic diffusion processes.
We could, hence, draw a promising line connecting our results to those developed within the set-up of large deviations theory. \\
This is a first promising step of an ambitious program, which definitely calls for further investigation.

\ack
The Authors wish to thank Christian Maes, Lamberto Rondoni and David Ruelle for inspiring discussions and clarifying remarks, and J. Wouters for useful comments.


\section*{References}
\begin{thebibliography}{50}


\bibitem{Kubo2} R. Kubo, Statistical-mechanical theory of irreversible processes: I. General theory and simple applications to magnetic and conduction problems
\textit{J. Phys. Soc. Japan} {\bf 12} 570 (1957).

\bibitem{Kubo} R. Kubo, The fluctuation-dissipation theorem, \textit{Rep. Prog. Phys.} \textbf{29}, 255 (1966).

\bibitem{BPRV} U. Marini Bettolo Marconi, A. Puglisi, L. Rondoni, A. Vulpiani, Fluctuation-Dissipation: Response Theory in Statistical Physics,
\textit{Phys. Rep.} \textbf{461}, 111 (2008).

\bibitem{Zub2} D. N. Zubarev, Nonequilibrium Statistical Thermodynamics, \textit{Consultant Bureau, New York} (1974).

\bibitem{Falc} M. Falcioni, A. Vulpiani, The relevance of chaos for the linear response theory, \textit{Phys. A} \textbf{215}, 481 (1995).

\bibitem{R3} D. Ruelle, General linear response formula in statistical mechanics, and the fluctuation-dissipation theorem far from equilibrium, \textit{Phys. Lett. A} \textbf{245}, 220 (1998).

\bibitem{Grits}  A. Gritsun, G. Branstator, Climate Response using a three-dimensional operator based on the fluctuation-dissipation theorem, \textit{Journal of the Atmospheric Sciences} \textbf{64}, 2558 (2007).

\bibitem{Grits2} A. Gritsun, G. Branstator, and A. Majda, Climate response of linear and quadratic functionals using the fluctuation dissipation theorem, \textit{J. Atmos. Sci.}, \textbf{65}, 2824 (2008).

\bibitem{Abra} R. Abramov, A. Majda, Blended response algorithms for linear fluctuation-dissipation for complex nonlinear dynamical systems, \textit{Nonlinearity}, \textbf{20}, 2793 (2007).

\bibitem{Zwanzig} R. Zwanzig, Nonequilibrium Statistical Mechanics, \textit{Oxford University Press} (2001).

\bibitem{Wou} J. Wouters, V. Lucarini, Disentangling multi-level systems: averaging, correlations and memory, \textit{J. Stat. Mech.} (2012) P03003.

\bibitem{Wou2} J. Wouters, and V. Lucarini, Multi-level dynamical systems: Connecting the Ruelle response theory and the Mori-Zwanzig approach, \textit{J. Stat. Phys.}, \textbf{151}, 850 (2013).

\bibitem{CRV}  M. Colangeli, L. Rondoni, A. Vulpiani, Fluctuation-dissipation relation for chaotic non-Hamiltonian systems, \textit{J. Stat. Mech.} (2012) L04002.

\bibitem{CMW} M. Colangeli, C. Maes, B. Wynants, A meaningful expansion around detailed balance, \textit{J. Phys. A: Math. and Theor.} \textbf{44}, 095001 (2011).

\bibitem{Evans1} D. J. Evans, D. J. Searles, S. R. Williams, On the Fluctuation Theorem for the Dissipation Function and its connection with Response Theory, \textit{J. Chem. Phys.} \textbf{128}, 014504 (2008).

\bibitem{MN} C. Maes, K. Neto\v{c}n{\'y}, Rigorous meaning of McLennan ensembles, {\it J. Math. Phys.} {\bf 51}, 015219 (2010).

\bibitem{R1} D. Ruelle, Smooth Dynamics and New Theoretical Ideas in Nonequilibrium Statistical Mechanics, \textit{J. Stat. Phys.} \textbf{95}, 393 (1999).

\bibitem{R2} D. Ruelle, Gaps and New Ideas in our Understanding of Nonequilibrium, \textit{Physica A} \textbf{263}, 540 (1999).

\bibitem{LC} V. Lucarini, M. Colangeli, Beyond the linear fluctuation-dissipation theorem: the role of causality, \textit{J. Stat. Mech.} (2012) P05013.

\bibitem{Lucar} V. Lucarini V, J. J. Saarinen, K.-E. Peiponen, E. M. Vartiainen,  Kramers-Kronig Relations in Optical
Materials Research (Heidelberg, Springer, 2005).

\bibitem{Agarw} G. S. Agarwal, Fluctuation-dissipation theorems for systems in non-thermal equilibrium and applications, \textit{Z. Phys.} \textbf{252}, 25 (1972).

\bibitem{Baies} M. Baiesi, C. Maes, An update on the nonequilibrium linear response, \textit{New Journal of Physics} \textbf{15} 013004 (2013).

\bibitem{Weid} W. Weidlich, Fluctuation-dissipation theorem for a class of stationary open systems \textit{Z. Phys.} \textbf{248} 234 (1971).

\bibitem{Boch1} G. N. Bochkov, Yu. E. Kuzolev, Nonlinear fluctuation-dissipation relations and stochastic models in nonequilibrium thermodynamics: I. Generalized fluctuation-dissipation theorem, \textit{Phys. A} \textbf{106}, 443 (1981).

\bibitem{Boch2} G. N. Bochkov, Yu. E. Kuzolev, Fluctuation-dissipation relations: achievements and misunderstandings, \textit{Phys. Usp.} \textbf{56}, 6 (2013).

\bibitem{Hang} P. H\"{a}nggi, H. Thomas, Stochastic processes: time-evolution, symmetries and linear response, \textit{Phys. Rep.} \textbf{88}, 207 (1982).

\bibitem{Evans2} D. J. Searles, L. Rondoni, D. J. Evans, The Steady state Fluctuation Relation for the Dissipation Function, \textit{J. Stat. Phys.} \textbf{128}, 1337 (2007).

\bibitem{McL} J. A. McLennan Jr., Statistical mechanics of the steady state, \textit{Phys. Rev.} \textbf{115}, 1405 (1959).

\bibitem{Zub} D. N. Zubarev, V. P. Kalashnikov, Extremal properties of the nonequilibrium statistical operator, \textit{Theor. Math. Phys.} \textbf{1} 108 (1969).

\bibitem{Maes1} M. Baiesi, C. Maes, B. Wynants, Nonequilibrium linear response for Markov dynamics, I: jump processes and overdamped diffusions, \textit{J. Stat. Phys.} \textbf{137}, 1094 (2009).

\bibitem{RQ} J. A. G. Roberts, G. R. W. Quispel, Chaos and time-reversal symmetry. Order and chaos in reversible dynamical systems,
\textit{Phys. Rep.} {\bf 216}, 63 (1992).

\bibitem{CK} M. Colangeli, P. De Gregorio, R. Klages, L. Rondoni, Steady state fluctuation relation with discontinuous ``time reversibility'' and invariant measures,
\textit{J. Stat. Mech.} (2011) P04021.

\bibitem{CR} M. Colangeli, L. Rondoni, Equilibrium, fluctuation relations and transport for irreversible deterministic dynamics,
\textit{Physica D} {\bf 241}, 6 (2012).

\bibitem{BLV} G.~Boffetta, G.~Lacorata,S.~Musacchio, A.~Vulpiani, Relaxation of finite perturbations: beyond the fluctuation dissipation relation,\textit{Chaos} \textbf{13}, 3 (2003).

\bibitem{Leb} F. Bonetto, A. Kupiainen, J.L. Lebowitz, Absolute continuity of projected SRB measures of coupled Arnold cat
map lattices, \textit{Ergod. Th. Dynam. Sys.} {\bf 25}, 59 (2005).

\bibitem{R4} D. Ruelle, Nonequilibrium statistical mechanics near equilibrium: computing higher order terms, \textit{Nonlinearity} \textbf{11}, 5 (1998).

\bibitem{Young} W. Cowieson, L.S. Young, SRB measures as zero-noise limits, \textit{Ergodic Theory and Dynamical Systems} \textbf{25}, 1115 (2005).

\bibitem{Kifer} Yu. Kifer, \textit{Random perturbations of dynamical systems} (Birkh\"{a}user, Boston, 1988).

\bibitem{MNW} C. Maes, K. Neto\v{c}n{\'y}, B.~Wynants Steady state statistics of driven diffusions, \textit{Physica A} {\bf 387}, 2675 (2008).

\bibitem{risken} H. Risken, \textit{The Fokker-Planck Equation}, 2nd edn.  (Springer Berlin, 1989).

\bibitem{dembo} A. Dembo, O. Zeitouni, Large Deviations Techniques and Applications (Springer-Verlag, New York, 1998).

\bibitem{poincare} C. Maes, On the origin and the use of fluctuation relations for the entropy, {\it S\'eminaire Poincar\'e} {\bf 2}, 29 (2003).

\bibitem{Katz} S. Katz, J. L. Lebowitz, and H. Spohn, Phase transitions in stationary nonequilibrium states of model lattice systems, \textit{Phys. Rev. B}, \textbf{28} 1655 (1983).

\bibitem{Prost} J. Prost, J. F. Joanny, J. M. Parrondo, Generalized fluctuation-dissipation theorem for steady-state systems, \textit{Phys. Rev. Lett.} \textbf{103}, 090601 (2009).

\bibitem{Gaw} R. Chetrite, K. Gaw\c{e}dzki, Eulerian and Lagrangian pictures of non-equilibrium diffusions, \textit{J. Stat. Phys.} \textbf{137}, 890 (2009).

\bibitem{Maes2} M.~Baiesi, E.~Boksenbojm, C.~Maes, B.~Wynants, Nonequilibrium Linear Response for Markov Dynamics,  II: Inertial Dynamics, {\it J. Stat. Phys.} {\bf 139}, 492 (2010).

\end{thebibliography}
\end{document}